\documentclass[12pt]{spieman}
\usepackage{amsmath, amssymb}
\usepackage{graphicx}
\usepackage{color}
\usepackage{subcaption}
\usepackage{setspace}
\usepackage{tocloft}

\newcommand{\mycitet}[1]{Ref.~\citenum{#1}}
\newcommand{\mycitep}[1]{\cite{#1}}

\title{Demonstration of high contrast with an obscured aperture with the WFIRST-AFTA shaped pupil coronagraph}
\author{Eric Cady\supscrsm{a}, Camilo Mejia Prada\supscrsm{a}, Xin An\supscrsm{a}, Kunjithapatham Balasubramanian\supscrsm{a}, \\
Rosemary Diaz\supscrsm{a}, 
  N. Jeremy Kasdin\supscrsm{b}, Brian Kern\supscrsm{a}, 
  Andreas Kuhnert\supscrsm{a}, Bijan Nemati\supscrsm{a},  \\
 Ilya Poberezhskiy\supscrsm{a},  A. J. Riggs\supscrsm{b},  Robert Zimmer\supscrsm{a}, Neil Zimmerman\supscrsm{b}}
\affiliation{\supscrsm{a}Jet Propulsion Laboratory, California Institute of Technology, Pasadena, CA, USA, 91109 \\
\supscrsm{b}Dept. of Mechanical and Aerospace Engineering, Princeton University, Princeton, NJ, USA, 08544
}

\cftpagenumbersoff{figure}
\cftpagenumbersoff{table} 
\begin{document}
\maketitle
\let\thefootnote\relax\footnote{\copyright 2015 California Institute of Technology. Government sponsorship acknowledged.}

\begin{abstract}
The coronagraph instrument on the WFIRST-AFTA mission study has two coronagraphic architectures, shaped pupil and hybrid Lyot, which may be interchanged for use in different observing scenarios.  Each architecture relies on newly-developed mask components to function in the presence of the AFTA aperture, and so both must be matured to a high technology readiness level (TRL) in advance of the mission.  A series of milestones were set to track the development of the technologies required for the instrument; in this paper, we report on completion of WFIRST-AFTA Coronagraph Milestone 2---a narrowband $10^{-8}$ contrast test with static aberrations for the shaped pupil---and the plans for the upcoming broadband Coronagraph Milestone 5.
\end{abstract}


{\noindent \footnotesize{\bf Address all correspondence to}: Eric Cady, Jet Propulsion Laboratory, 4800 Oak Grove Drive, Pasadena, CA, USA, 91101; Tel: +1 818-393-8254
Email: \linkable{eric.j.cady@jpl.nasa.gov}}

\begin{spacing}{2}   

\section{Introduction}

The proposed NASA WFIRST-AFTA observatory consists of a 2.4m telescope, gifted from another agency, coupled to a wide-field infrared instrument, whose task is to address a variety of top-level science goals laid out in the Astro2010 Decadal Survey \mycitep{Spe15, NRC10}  (WFIRST: Wide Field InfraRed Survey Telescope, AFTA: Astrophysics-Focused Telescope Assets).  This aperture is larger than the 1.5m aperture proposed in the Decadal Survey for a wide-field mission, improving the angular resolution, and a second instrument, a coronagraph, was designed to capitalize on this for exoplanet science, and also to address technology development goals for high-contrast imaging discussed in the Decadal Survey.

A subsequent report \mycitep{NRC14} found that adding a coronagraph, which was less technologically mature than the wide-field instrument, increased the risk of cost growth, but was not fully mature enough to evaluate adequately.  As a result they recommended, and NASA implemented, a program to ``aggressively mature the coronagraph design'' \mycitep{NRC14} as well as required supporting technology in advance of an independent review of the instrument. 

A series of 9 milestones was developed to demonstrate the technology readiness of the coronagraphic instrument under an aggressive schedule.  This instrument is baselined to use two separate but compatible modes, the shaped pupil coronagraph (SPC)\mycitep{Car13, Rig14} and the hybrid Lyot coronagraph (HLC)\mycitep{Tra13}.  Another design is based around a phase-induced amplitude apodization complex-mask coronagraph (PIAACMC)\mycitep{Guy14}, and it is being matured as a backup.  In addition, a low-order wavefront sensor is being developed to track and compensate for slow thermal drifts and line-of-sight jitter in the upstream optical train\mycitep{Pob14}.  These milestones cover all four of these technologies, as well as the instrument detectors, and are reviewed by an independent Technical Analysis Committee (TAC).  Successful completion of all nine will bring the instrument to NASA Technology Readiness Level 5 (TRL 5) in September of 2016.  Four of these milestones relate directly to the shaped pupil (Milestones 1, 2, 5, and 9).  The completion of WFIRST-AFTA Milestone 2, and preparations for Milestone 5, will be the subject of this paper.

As laid out in original definitions to NASA Headquarters and the TAC, WFIRST-AFTA Coronagraph Milestone 2 was defined as:
\begin{center}
\em{The shaped pupil coronagraph in the High Contrast Imaging Testbed demonstrates $10^{-8}$ raw contrast with narrowband light at 550 nm in a static environment.}
\end{center}

This milestone was successful; we achieved repeated convergence to a $\sim 6\times 10^{-9}$ mean contrast in the High Contrast Imaging Testbed (HCIT), across a $4.4-11.2 \lambda/D$ wedge-shaped dark hole with a shaped-pupil characterization mask in a $2\%$ band centered around 550nm.  (More precise contrasts and error bars are given later in Table \ref{table:contdata}.)  These materials were due September 30, 2014 to the TAC; milestone materials were presented to the TAC September 17th and reviewed with the TAC October 8th.  These materials were accepted, and the milestone completion was endorsed.  The next milestone for the shaped pupil coronagraph, Milestone 5, will be in September 2015, and will require demonstrating equivalent contrast in a $10\%$ band with a newer shaped pupil design.  (A third contrast milestone in September 2016, Milestone 9, will have the same contrast demonstration requirements as Milestone 5, but in the presence of dynamic wavefront errors introduced by a telescope simulator and actively corrected by a low-order wavefront sensing and control subsystem.)

Section \ref{sec:bg} will lay out the technical background behind the development of the shaped pupil coronagraph and its mask designs, Sect. \ref{sec:thp} will give the theoretical performance expectations, and Sect. \ref{sec:fed} will detail the hardware and software used for the testing.  These results and their associated analysis, including comparisons with models and estimates of planet yield given the results seen in the testbed, are presented in Sect. \ref{sec:exp}.  Section \ref{sec:fut} will discuss upcoming testing for Milestone 5, as well as longer-term plans and related activities.

\section{Background} \label{sec:bg}

\begin{figure}
\centering
\includegraphics[width=0.75\textwidth]{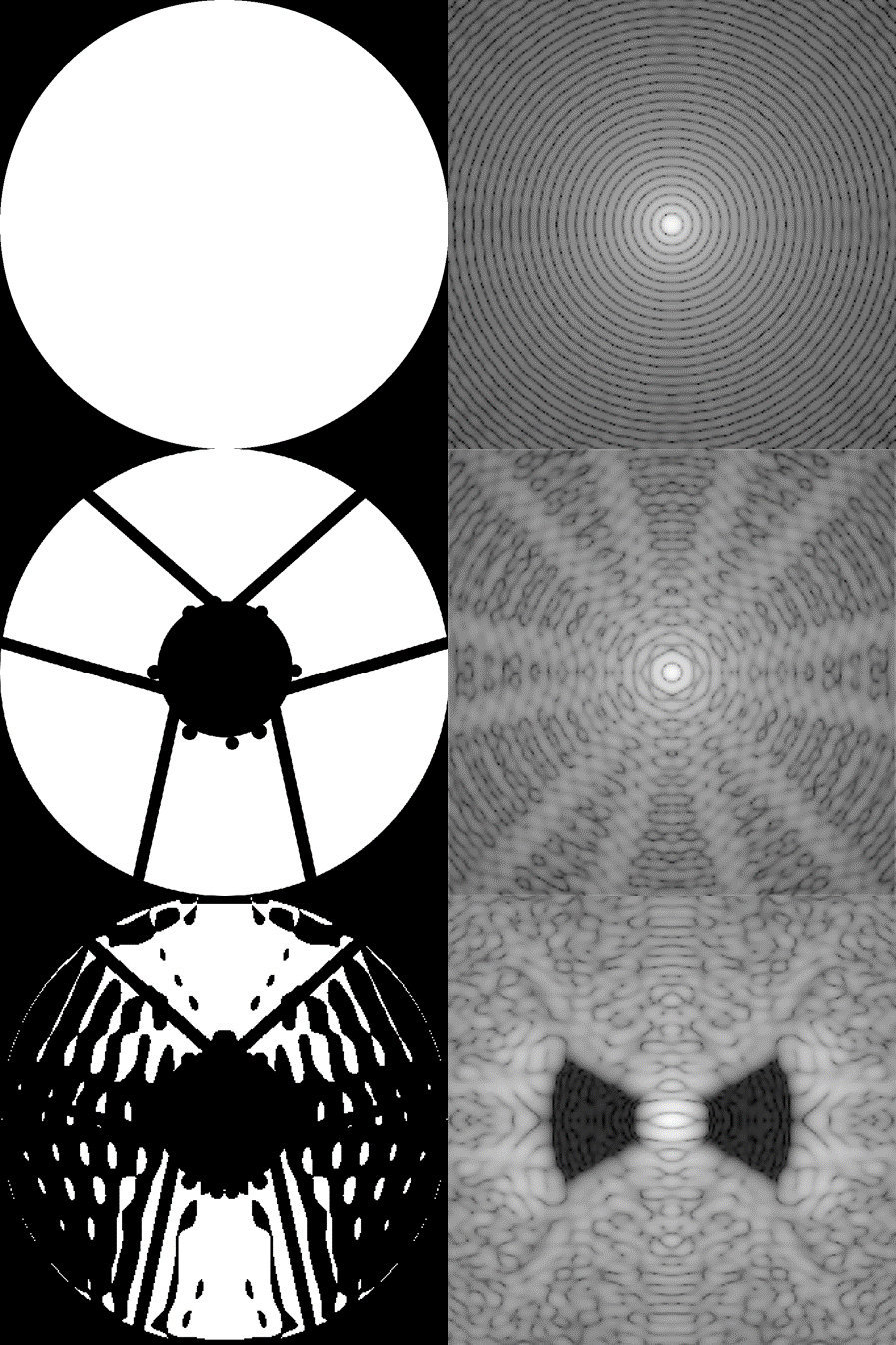}
\caption{\textit{Top:} In its focal plane, an unobstructed circular aperture produces an Airy pattern: a circularly-symmetric point spread function whose falloff with radius is too slow to permit the detection of nearby planets. \textit{Middle:} The AFTA pupil is further complicated by a secondary obscuration, spiders, and some features on the primary mirror, which raise the wings of the point spread function.  \textit{Bottom:} In its focal plane, a shaped pupil creates regions of high contrast where planets may be observed, even with the shape of the AFTA pupil embedded.  All focal plane images are on the same logarithmic scale.} \label{fig:whatsp}
\end{figure}

The concept of the shaped pupil coronagraph is based in the principles of Fourier optics: under certain assumptions (e.g. paraxial beam), the relationship between the electric fields at a pupil plane and at a focal plane in an optical system can be well represented by a Fourier transform.  Given a circular aperture, for example, the wavefront from a point source will be focused into an amplitude distribution $\propto J_{1}(x)/x$, giving rise to the well-known Airy pattern in its point spread function (PSF).  With a shaped pupil, that circular aperture is replaced by a shape---generally chosen by an optimization process\mycitep{Van03, Van04, Car13, Rig14, Zim15}---whose Fourier transform has regions where the amplitude of the fields is extremely low, producing a high contrast between the peak of the PSF and these regions in the wings.  Should a planet be present off-axis, it will be detectable in these regions.  (Figure \ref{fig:whatsp} gives an example of this relationship.)

At its simplest, a shaped stop in a pupil plane is the entirety of the coronagraphic optics in the SPC, barring the addition of a field stop to keep the camera from saturating from the high dynamic range involved. This makes the shaped pupil both straightforward to implement and quite robust to low-order errors, as the stellar PSF does not have to be exactly aligned behind a diffractive focal plane mask (FPM).  However, the downside of this simplicity is reduced science yield: the IWA is generally larger than for other coronagraphs with open apertures, and the throughput is also often lower because of a spread-out PSF core and the additional pupil obscurations.  Other concepts, such as the hybrid Lyot coronagraph or PIAACMC, are expected to have the capability to have deeper contrasts and work closer to the star.  

The shaped pupil Lyot coronagraph (SPLC)\mycitep{Rig14, Zim15} was created to try improve science capability without sacrificing too much of the robustness of the shaped pupil; it will use the field stop actively as a focal plane mask and augment this with a Lyot stop at a downstream pupil.  Most of the contrast is still generated by the SP apodization; only $10^{3}$ of the $10^{9}$ suppression factor is due to the focal plane mask and Lyot stop.  This development was not finished by Milestone 2, but an SPLC will be used as a part of the next shaped pupil contrast milestone (WFIRST-AFTA Coronagraph Milestone 5, scheduled for September 2015).  

\begin{figure}
\centering
\includegraphics[width=1\textwidth]{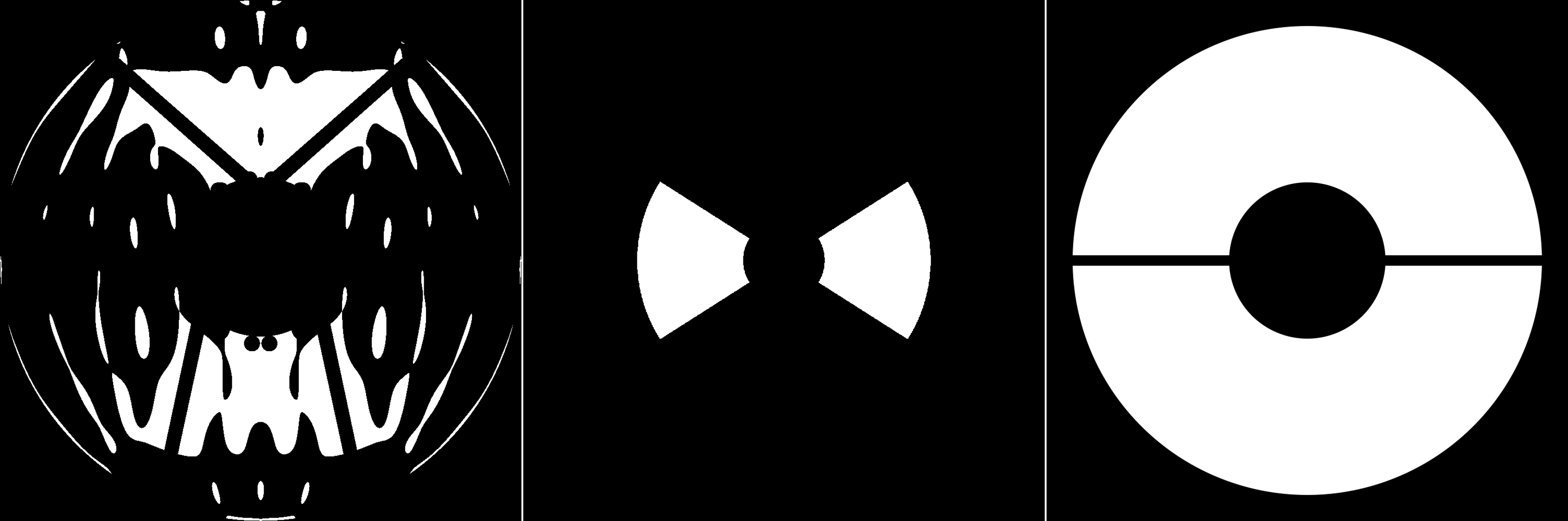}
\caption{The three masks of a shaped pupil Lyot coronagraph.  \textit{Left.} Shaped pupil. \textit{Center.} Bowtie focal plane mask. \textit{Right.} Lyot stop.}
\label{fig:splc}
\end{figure}

\begin{figure}
\centering
\includegraphics[width=0.5\textwidth]{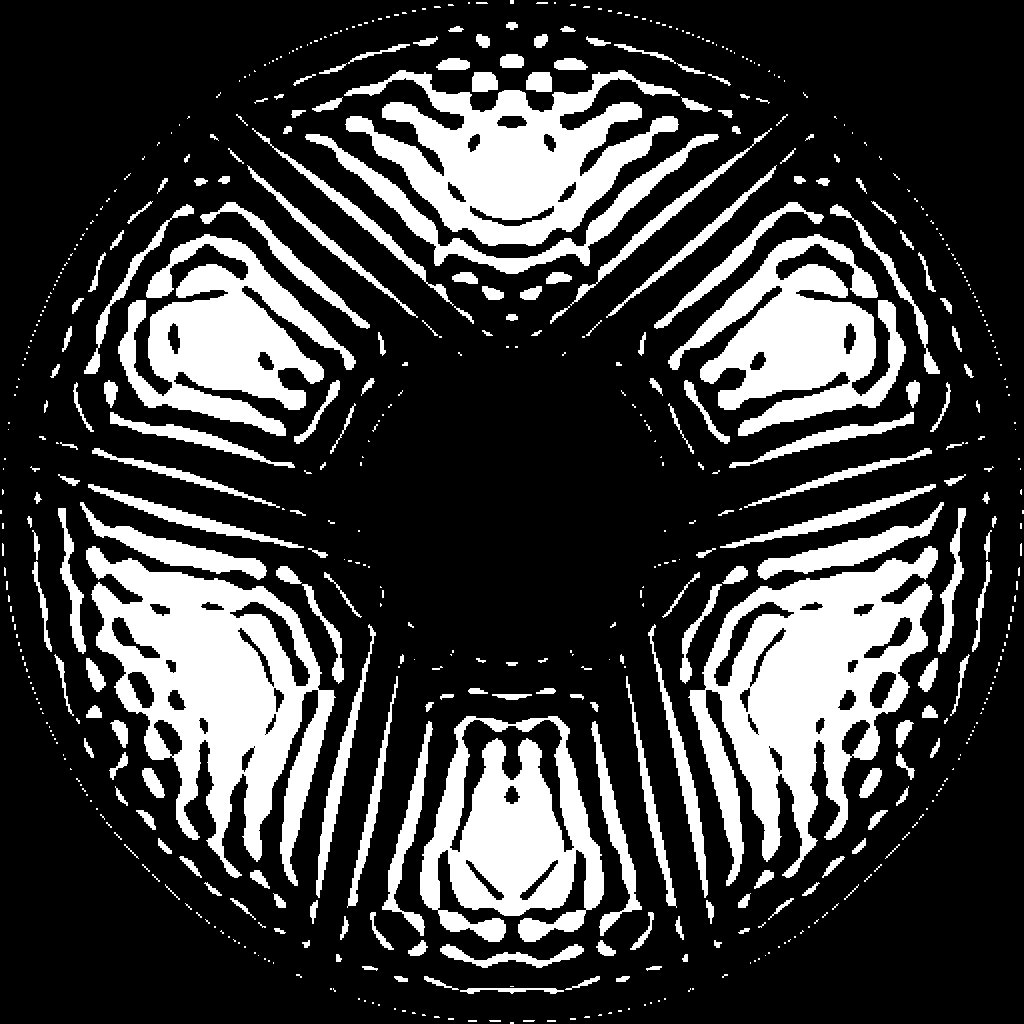}
\caption{A disk mask, with an inner working angle of $6\lambda/D$ and outer working angle of $24 \lambda/D$.  A second-generation mask to be used with a Lyot stop is under development.}
\label{fig:disk}
\end{figure}

Two general classes of shaped-pupil mask are being optimized for WFIRST-AFTA: characterization and disk science.  Characterization masks use sets of wedged dark regions to improve their inner working angle (IWA), but require multiple masks to image the whole of the area around the star.  (The form of the field stops sized for shaped pupils with wedge-shaped regions of high suppression, such as in Fig. \ref{fig:whatsp}, has led them to be termed generally as ``bowtie masks''.)  To compensate for this on WFIRST-AFTA, they are expected to be used with the integral field spectrograph (IFS) to spectrally characterize planets whose focal-plane locations have already been determined.  Disk science masks have poorer inner working angle, but can create a dark region $360^{\circ}$ about the star for imaging of disk structures.  A first-generation disk science mask is shown in Fig. \ref{fig:disk}; an SPLC version of the disk science mask is still under development, and neither has been able to be tested yet.  For Milestone 2, a characterization mask was used with a dark hole running from $4.4-11.2\lambda/D$ in a $52^{\circ}$ wedge, covering an area of approximately $48.1$ $(\lambda/D)^2$.  (See \mycitet{Rig14} and \citenum{Zim15} for a discussion of the trade-offs in the choices of inner and outer working angles.) Subsequent design use larger wedge angles and are further detailed in Sect. \ref{subsec:masks}.

Shaped pupils in the past\mycitep{Van03, Van04, Bel07} have been transmissive, primarily using holes etched through wafers with deep reactive ion etching (DRIE) to create the pupil shape\mycitep{Bal06}.  These masks were freestanding and self-supporting, and were successfully demonstrated at high contrast in the HCIT previously, including $4\times 10^{-9}$ contrast in monochromatic, $790$nm laser light with 2DMs in the summer of 2013 \mycitep{Rig13}. Unfortunately, the features of the AFTA pupil---with a slightly-off-axis secondary, six spiders, and additional small circles around the secondary---proved hostile enough to optimization that pupil shapes could not be found without small isolated regions that could not be built in a freestanding optic, and new technology was required.  This was not the case for the bowties, and they have continued to be produced as etched apertures in wafers.

A few transmissive masks had been created on a transparent substrate for specific applications, such as apodized-pupil Lyot coronagraphs for ground-based telescopes, but the ghosts induced by internal reflections off the substrate surfaces have not made these designs suitable for contrasts at the levels required by WFIRST-AFTA.  Instead, the mask was made reflective, with nominally-open region coated with reflective metal and the dark region created from a highly-absorbing material: black silicon\mycitep{Bal13}.  The manufacturing of this mask and associated model validation was the subject of WFIRST-AFTA Coronagraph Milestone 1, and is covered in more detail in \mycitet{Pob14}. (Subsequent technology development has shown that substrates exist with sufficiently-low reflectivity that transmissive masks could be potentially be manufactured for very-high-contrast applications\mycitep{Hob07}, but they have yet to be qualified to the level of reflective designs.)

\section{Theoretical performance} \label{sec:thp}

\begin{figure}
\centering
\includegraphics[width=1\textwidth]{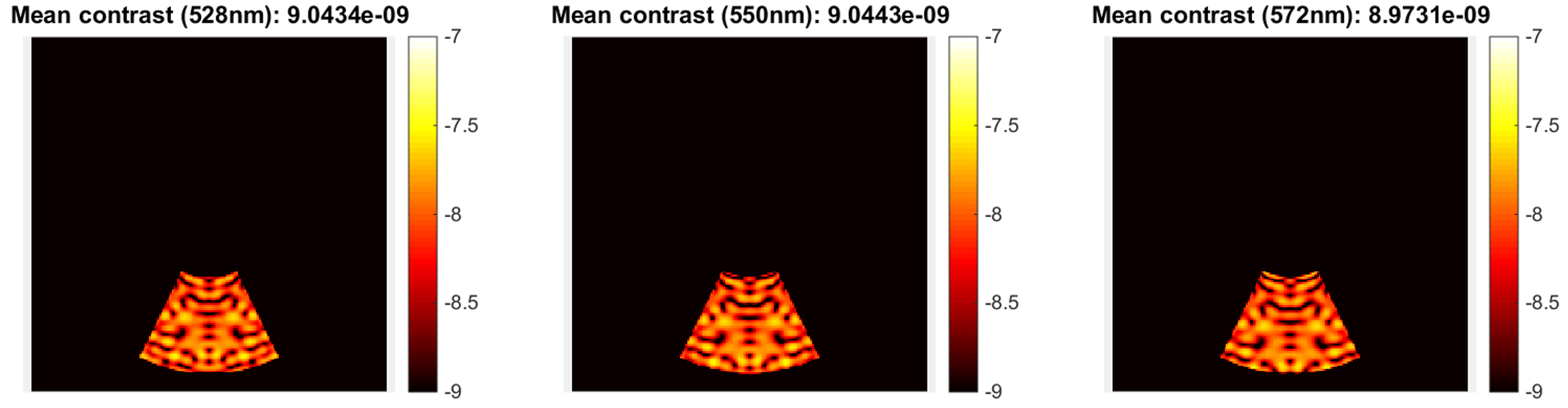}
\caption{Design performance of the first-generation shaped pupil at the central wavelength of each of 3 $2\%$ bands, representing the ends and center of a $10\%$ band.  Results are normalized such that the peak brightness of the PSF is 1 and displayed on a log$_{10}$ scale.  These results assume no aberrations, no deformable mirrors, and no active correction.}
\label{fig:ol5b}
\end{figure}

The first-generation shaped pupil for WFIRST-AFTA, with no Lyot stop, was meant to produce a contrast of just under $1\times 10^{-8}$ over a $10\%$ band, as shown in Fig. \ref{fig:ol5b}\mycitep{Pob14}.  The bowtie was undersized with respect to the full dark region of the PSF, to ensure broadband performance as the PSF scales linearly with wavelength.  This can be seen more clearly by comparing the subfigures of Fig. \ref{fig:ol5b}: the intensity residual is stretched radially but does not alter its morphology.  Only a single side of the image plane is shown, to match the single-sided bowtie used in the Milestone 2 tests.  (In the absence of aberrations, this coronagraph produces a double-sided dark hole, as in the bottom right of Fig. \ref{fig:whatsp}, but with a single deformable mirror to correct amplitude and phase aberrations, only one side may be corrected at a time.)

These simulated results, obtained purely with Fourier methods, assume neither the presence of aberrations nor noise, which degrade contrast, nor the presence of a deformable mirror system and control loop which can compensate for those aberrations and potentially enhance the achievable contrast.

Unfortunately, simple models of closed-loop coronagraphy with a shaped pupil provide little insight into the system performance.  Given a set of static errors and knowledge of the optical system, the coronagraph eventually corrects all the errors occurring at controllable spatial frequencies.  The rate of that correction in simulation is set by a regularization parameter, which in a real system prevents noise, disturbances, or unmodeled system dynamics from causing the correction to diverge.  

\begin{figure}
\centering
\includegraphics[width=0.9\textwidth]{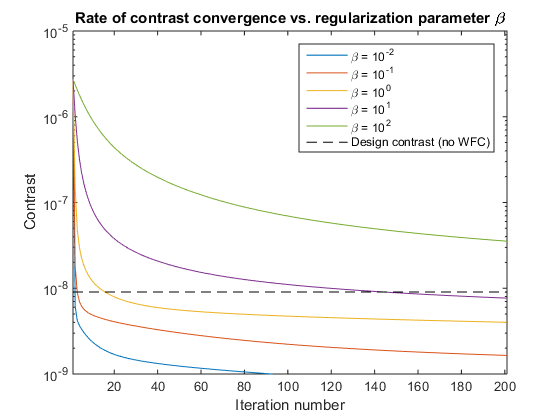}
\caption{
Simulated mean contrast as a function of iteration in a $4.4-11.2\lambda/D$ region at 550nm, using a model of the HCIT with phase aberrations and electric field conjugation (EFC) with one of several regularization parameters to correct them.  The dashed line shows the mean contrast in this region from base design, prior to the use of wavefront control.}
\label{fig:cl1b}
\end{figure}

Figure \ref{fig:cl1b} shows the contrast as a function of iteration number for a $2\%$ band and a sample of typical regularization parameters used in HCIT control schemes.  These parameters are the same ones used in testbed control as described later in Sect. \ref{subsec:sac}, and the control region is the same as shown in Fig. \ref{fig:ol5b}.  Each begins with the same simple set of static aberrations; all eventually outdo the nominal design, given sufficient time.  This is not a surprise, as DMs introduce an additional piece of leverage by providing phase (and some amplitude) control in the pupil, whereas the shaped pupil mask only gives amplitude control.

Details of the estimation and control algorithms in use are given briefly in Sect. \ref{subsec:sac} and at greater length in \mycitet{Giv11} and \mycitet{Giv09}.

\section{Facility and experiment design} \label{sec:fed}

\subsection{Masks} \label{subsec:masks}

\begin{figure}
\centering
\includegraphics[width=1\textwidth]{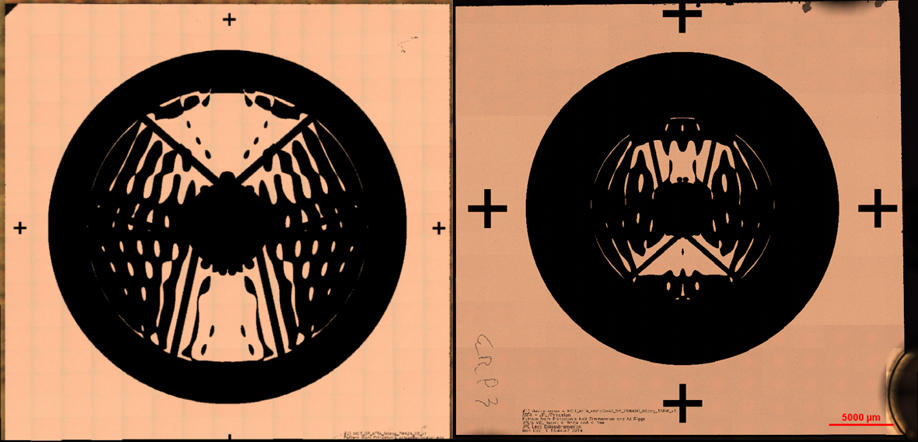}
\caption{\textit{Left.} A downsampled version of high-resolution microscope imagery of the Milestone 2 mask, taken as part of manufacturing validation.  The background lines are a stitching artifact.  \textit{Right.} A downsampled version of high-resolution microscope imagery of the Milestone 5 mask, taken as part of manufacturing validation.  The faint horizontal lines are a stitching artifact.} \label{fig:m25mask}
\end{figure}

The milestone 2 mask is a 35mm$\times$35mm diced section of a silicon wafer 2mm thick.  This mask is coated with bare aluminum, and black silicon is used to create the dark regions of the pupil.   An image of the mask is shown at left in Fig. \ref{fig:m25mask}.  Milestone 1 previously demonstrated that this mask, as manufactured, was sufficient to permit demonstrations at the contrast level required for Milestone 2.  (See \mycitet{Bal15} for a fuller discussion of the qualification and error analysis.)  The 2mm mask thickness was selected to help minimize wafer bow, which introduces undesirable low-order phase aberrations into the pupil; in addition, a pre-selection process was done to find wafers with minimal Zernike components above focus in the prospective shaped-pupil region.  Focus can be compensated for at a system level by translating the camera and field stop; this proves fortunate, as the focus component of the Milestone 2 mask introduces microns of wavefront, well outside the stroke capability of our DM.   Subsequent masks have been made using 4mm thick wafers, and the bow problem has been reduced considerably. The Milestone 5 SPLC mask, made on one of these wafers, is shown at right in Fig. \ref{fig:m25mask}.  All of these masks are achromatic, in that the aluminum and black silicon reflectivities change negligibly across the spatial extent of the pupil in our desired bandpasses.

\begin{figure}
\centering
\includegraphics[width=1\textwidth]{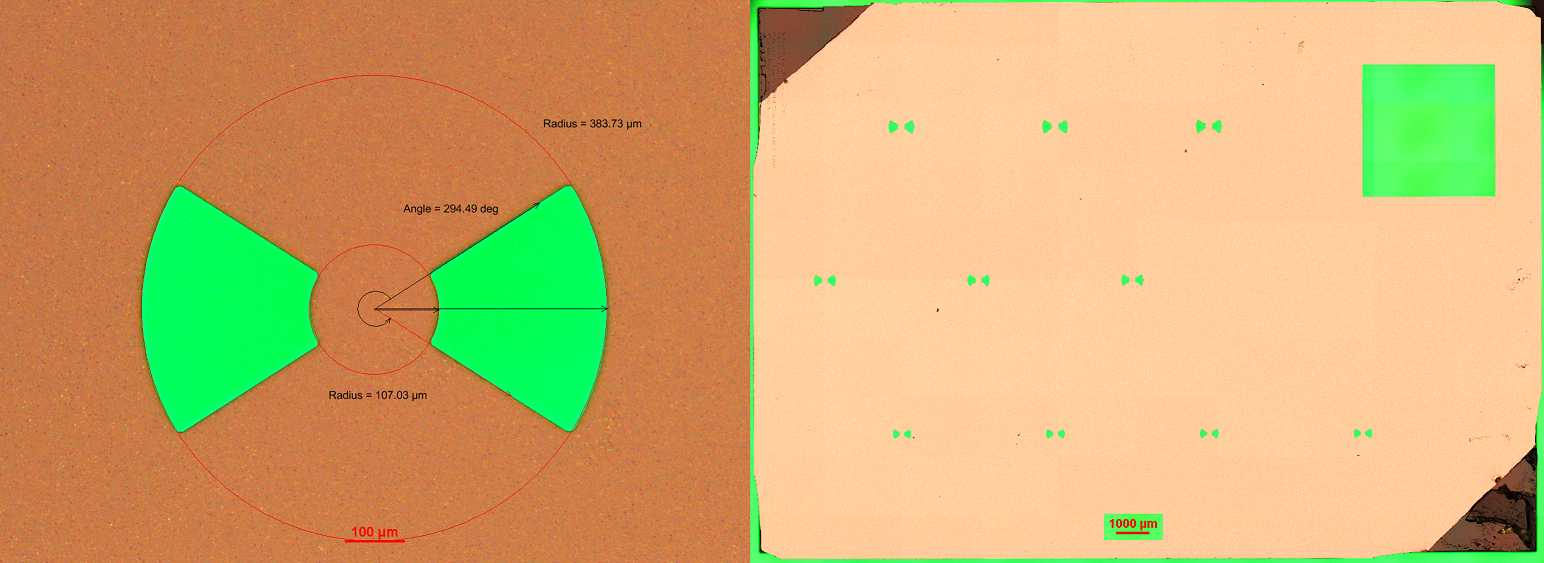}
\caption{\textit{Left.} A high-resolution microscope image, in transmission, of a single SPLC bowtie mask.  \textit{Right.} A high-resolution microscope image, in transmission, of an array of SPLC bowtie masks.  All are the same shape---$2.5-9\lambda/D$ over a $65^{\circ}$ wedge at a center wavelength, but that center wavelength is 770nm, 660nm, or 550nm respectively from top to bottom.  Also included are a pinhole in the second row and a open square to allow the field stop to be ``removed'' from the beam.} \label{fig:combfs}
\end{figure}

An array of field stops were etched into a 24mm $\times$ 16mm section of an SOI wafer.  Figure \ref{fig:combfs}, at left, shows the detail of one stop taken in transmission with a microscope as part of manufacturing validation.  The angular sizes of the bowties were undersized with respect to dark-hole wedge angle, which was $60^{\circ}$, to permit simpler alignment; newer SPLC designs use $65^{\circ}$ wedges on their dark holes, permitting three masks to cover the entire azimuthal range in the image plane with some overlap.  An example of one of these arrays in given in on the right in Fig. \ref{fig:combfs}.

In either case, these masks purely function in transmission; the front face is coated with metal and reflects the undesired light into a beam dump. Eventually, in Milestone 9, reflection from the bowties will instead have its PSF core phase-shifted and be redirected to the low-order wavefront sensor to track slowly-varying aberrations upstream of the coronagraphic mask.

\begin{figure}
\centering
\includegraphics[width=1\textwidth]{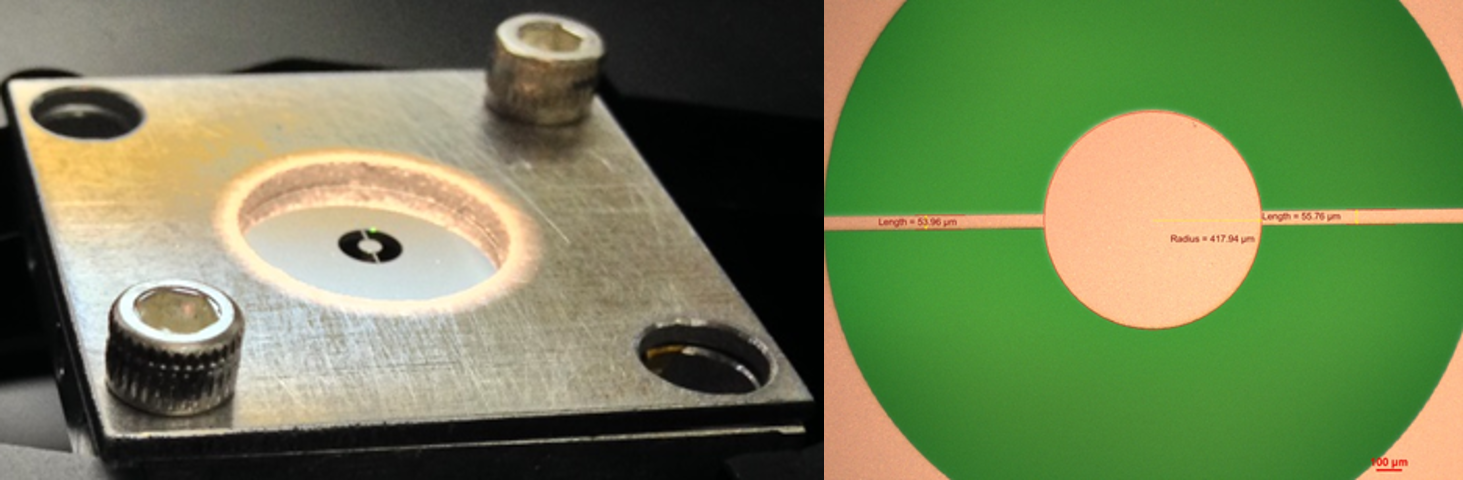}
\caption{\textit{Left.} An SPLC Lyot mask mounted in its holder in preparation for microscopy.  \textit{Right.} The same mask viewed in transmission and reflection; measured distances are marked.  The light source providing transmitting illumination is green.}\label{fig:lyot}
\end{figure}

The final mask in an SPLC is a Lyot stop, which is also etched with DRIE from an SOI wafer.  (One is neither present nor necessary in a baseline shaped pupil coronagraph.)  This stop blocks light diffracted from the edges of the field stop, and permits smaller inner working angles and deeper overall contrast than would be achievable otherwise.  Figure \ref{fig:lyot} shows the mask mounted in its holder and viewed in transmission and reflection in a microscope.  The thin features holding the large secondary are created by doing a partial etch over the region adjacent to the opening, so the cross-section of the thin struts have a T-shape with a high aspect ratio from strength while keeping thin edges.

\subsection{Optical layout and hardware}

The High Contrast Imaging Testbed (HCIT) at JPL is a NASA facility with a number of coronagraphic optical testbeds installed in vacuum chambers to eliminate the effects of air turbulence and simulate a space environment.  In the past, this has served as a facility where researchers could install and test their coronagraphs at higher contrasts than were accessible elsewhere, including shaped pupils in two instances \mycitep{Bel07, Rig13} prior to existence of the WFIRST-AFTA coronagraph instrument.  

\begin{figure}
\centering
\includegraphics[width=0.8\textwidth]{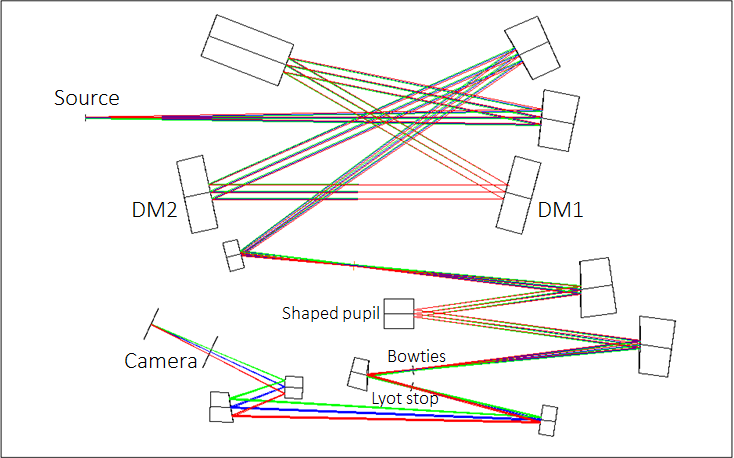}
\caption{A Zemax layout of the testbed in tank HCIT-1 during the Milestone 2 tests.  The text overlay points out salient hardware.  Unmarked optics are either flats or OAPs; the large rectangle represents a keep-out zone for a fine steering mirror that was ultimately not incorporated into this testbed.  A flat was used in its place.  The camera is mounted on a 10-inch stage, permitting it to reach both the pupil and focal planes indicated by black lines crossing the ray trace.  After the completion of Milestone 2 in October 2014, DM2 was installed; the Lyot stop was installed in April 2015.} \label{fig:layout}
\end{figure}

Initial optical alignment of bulk optics during testbed assembly is done with a Faro coordinate-measuring arm and an interferometer working in double-pass, working to mirror positions set by a Zemax layout of the system.  Figure \ref{fig:layout} shows this optical layout of the SPC testbed as designed; this will not be the layout for the final flight instrument.  Subsequent manual alignments of coronagraph masks are done with attached micrometers and fine-pitch knobs, and fine adjustments are performed in vacuum with remotely-controlled actuators.  Both use the science camera for feedback.  The testbed is fed from a single-mode fiber and can use any fiber-coupled source; currently we use an NKT Photonics EXB-6 with a VARIA tunable filter to provide our narrow and broad bandpasses for milestone goals, and a higher-powered 516nm laser as an alignment and testing source.

In addition to the basic requirements of the shaped pupil architecture (reflective shaped pupil, bowtie mask for dynamic range, at least 1 DM), two other considerations drove the final layout:
\begin{itemize}
\item \textbf{Prompt execution:} The shaped pupil contrast milestone was the first testbed-demonstration milestone for the WFIRST-AFTA coronagraph.  In order to get the testbed ready early to meet this milestone, the layout was designed to reuse existing optics as much as possible, including recycling all of the off-axis parabolas currently in use on the HCIT bench, and so avoiding lengthy procurements.   This decision also fixed the focal lengths in the system, necessitating additional folds to keep it confined to the optical table.  
\item \textbf{Future expansion:} While the initial layout only included a single DM and a reflective shaped pupil, it was designed to be compatible with the full extent of the combined SPC/HLC architecture.  Surrogate folds were added so that a second DM and a fast steering mirror could be swapped into those locations later, and a focal plane prior to the shaped pupil was kept free so a hybrid Lyot coronagraph could be used there, along with a reflective Lyot stop.  Despite the eventual plan to use two DMs, only one was available when the testbed was first assembled.  The shaped pupil does not require 2 DMs to create high contrast in a half-plane region, and Milestone 2 was not specified to require 2DM control for completion.  (A DM was later acquired from another testbed and integrated successfully into SPC in anticipation of Milestone 5 requirements.)
\end{itemize}

\subsection{Software, algorithms, and calibration} \label{subsec:sac}

Testbed control during experiments was performed using existing HCIT software for remote control of actuators, DMs, light sources, and the camera.  This software implements two loops which run concurrently: one interfaces at a low-level with the testbed hardware, putting probes on the DM and taking images, while the other does the high-level wavefront estimation and correction.  Interfaces between the loops are handled by passing of data files containing either desired DM commands or photometrically-corrected camera images.

Photometry is determined empirically by inserting pairs of speckles into the image plane with the bowtie removed, using sinusoidal patterns on the DM, and evaluating the relative flux between these speckles and the center of the PSF.  These speckles are chosen to be sufficiently large such that other aberrations do not affect the photometry.  A series of images is averaged to obtain this ratio, to minimize the effects of photon noise and source flux instability (measured to be less than $1\%$).   When the bowtie is replaced, these decentered speckles may be measured directly---since they pass through the holes in the bowtie---to determine detector counts per second in the PSF core, using the ratio determined previously.  The photometry may be recalculated at subsequent times by reapplying these DM settings without needing to remove the bowtie.

Plate-scale is checked with two independent methods, both of which have been shown to be self-consistent by internal testing.  The first uses the photometry speckles, which were applied with a known spatial frequency across the DM, and the plate scale is derivable directly from their spacing.  The second correlates the empirical PSF to a model of the PSF, comparing the locations of wing features without applying any DM shapes. 

Wavefront estimation is done with a pairwise estimation scheme, in which ``probes'' are placed on the DM to modulate the electric field across the region of interest.  For our purposes, these probes are made from a combination of sinc and sine functions chosen such that the probe amplitude over the region of interest is relatively uniform.  Two sinc functions on a pupil-plane DM in orthogonal Cartesian directions create a rectangular region of modulation in the image plane; multiplying this by a sinusoid in the pupil changes the spacing and the modulation phase \mycitep{Giv07}.

Given two or more pairs of probes, along with a images with no probes at all, we can back out both the complex electric field of the residual simulated starlight and the portion of the field that does not interact with the probes and hence is unlikely to be correctable.  (The components that do and do not interact with the probes are referred to as the ``coherent'' and ``incoherent'' parts, respectively.)  For further detail of the algorithm, see \mycitet{Giv11}.

Wavefront correction is done with the electric field conjugation (EFC) algorithm, which chooses DM settings to minimize the coherent part of the electric field across the dark hole, using a model of the coronagraph and DM.  A regularization parameter is included to weigh the relative importance of the data versus the model, and this parameter is tuned by taking a series of extra images per iteration.  For more detail on EFC, see \mycitet{Giv09}.

\section{Data and analysis} \label{sec:exp}

\subsection{Contrast results}

For the milestone demonstration, we collected three independent sets of data between 8/28/2014 and 8/31/2014, each attempting to minimize the residual light across the dark hole, which was chosen to match the size of the bowtie: $4.4-11.2 \lambda/D$.  For consistency, each run started from a flat-DM setting obtained earlier during testbed calibration.  Using three independent runs starting from flat has been a standard method of demonstrating coronagraph capability during previous demonstrations under the Technology Demonstration for Exoplanet Missions (TDEM) program, and the practice was continued for the WFIRST-AFTA coronagraph milestones, although starting from a previously-good DM setting produces much faster convergence, in 1-2 iterations in the absence of major drifts.  

\begin{figure}
\centering
\includegraphics[width=1\textwidth]{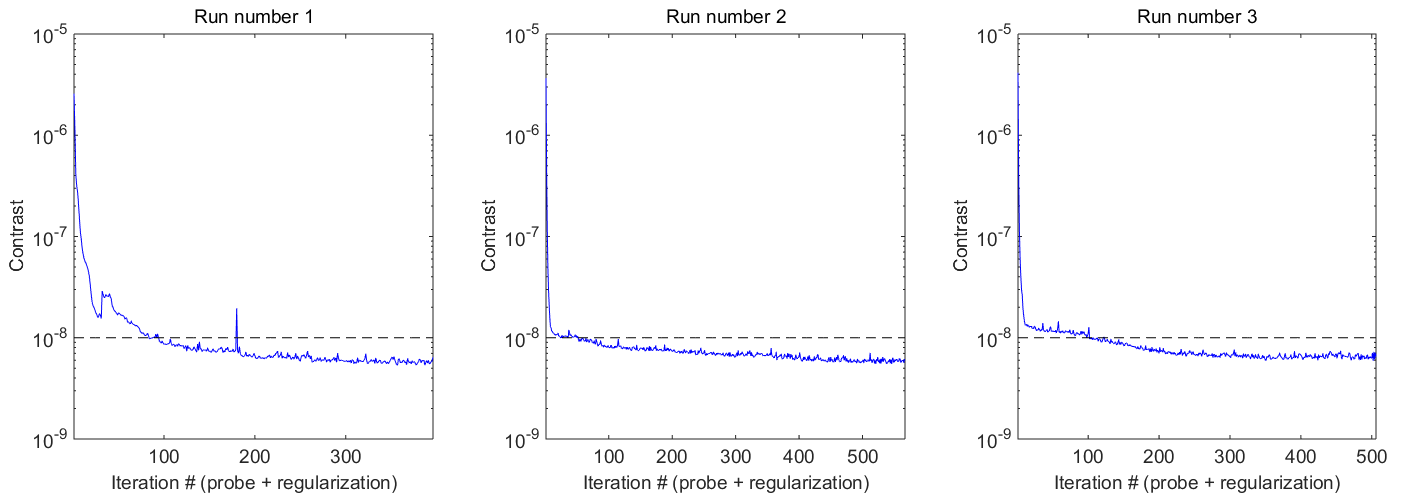}
\caption{Mean contrast across the dark hole as a function of iteration number, for each of the three Milestone 2 runs. Run 1 contained manual interventions while testing appropriate parameters for the run, producing the plateau near iterations 30-50 and the spike around iteration 180.} \label{fig:iters}
\end{figure}

Each run was allowed to continue until the contrast level had ceased to change; Figure \ref{fig:iters} shows the progression of contrast over time.  In each case, the contrast variation over the final 100 iterations was small, and Table \ref{table:contdata} gives  1) the estimate for the mean contrast across the dark hole over the final 100 iterations, 2) a $99\%$ confidence interval on the mean contrast achieved in the dark hole, and 3) the rms level of the DM correction applied from the beginning to the end of the run.  The variance used to compute the confidence interval is derived both from the temporal variance from iteration to iteration and the uncertainty in the photometric calibration procedure laid out in Sect. \ref{subsec:sac}.  Both 1 and 2 show the contrast level is significantly below the milestone threshold contrast of $1\times 10^{-8}$, and is repeatable from run to run.

\begin{table}[h] \caption{Contrast data for 3 Milestone 2 runs} \label{table:contdata}
\centering
\scriptsize{
\begin{tabular}{|c|ccc|}
\hline
Run & \#1 (8/28-8/29) & \#2 (8/29-8/30) & \#3 (8/30-8/31)  \\
\hline
Contrast over final 100 iterations & $(5.85\pm 0.49)\times 10^{-9}$ & $(5.95\pm 0.49)\times 10^{-9}$ & $(6.59\pm 0.56)\times 10^{-9}$ \\
$99\%$ CI on mean contrast in dark hole & $[5.54\times 10^{-9}, 6.16\times 10^{-9}]$ & $[5.64\times 10^{-9}, 6.26\times 10^{-9}]$ & $[6.23\times 10^{-9}, 6.94\times 10^{-9}]$ \\
magnitude of DM correction, rms & $1.46$nm & $1.66$nm & $1.77$nm  \\
\hline
\end{tabular}
}
\end{table}

\begin{figure}
\centering
\includegraphics[width=1\textwidth]{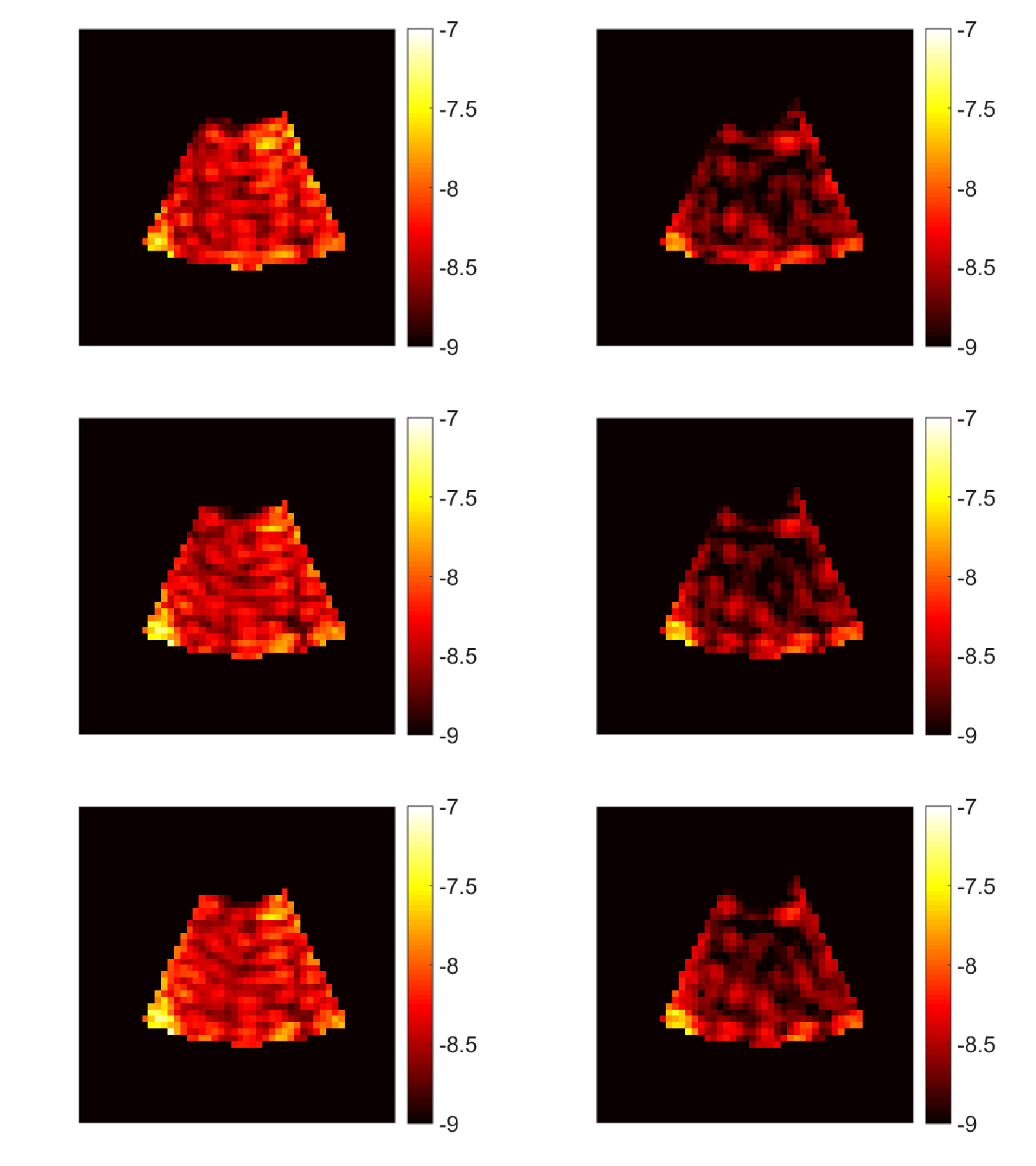}
\caption{\textit{Left:} Average light distribution across the dark hole for each of the three runs.  Area outside the dark hole is masked in software.  \textit{Right:}  The incoherent part (the portion of the intensity that cannot be modulated by the probes) for each of the three runs.  The color scale in all images is identical and is in units of contrast, on a log$_{10}$ scale.} \label{fig:meaninc100}
\end{figure}

The contrast is not uniformly distributed over the dark hole: the upper right and lower left corners of the dark hole are particularly bright, with large incoherent components, and these drive the estimates of contrast as a function of radius.
See Figure \ref{fig:meaninc100} for the contrast distribution across the dark hole and its incoherent parts, and Figure \ref{fig:contrad} for curves of contrast versus radius for the three runs.

One factor limiting the achieved contrast was a persistent noise on the DM voltage lines which caused the speckle field to ``boil'' despite the lack of any DM commands.    This was eventually traced to two issues---firmware on a driver card of an incorrect version, and a ground loop at the tens-of-millivolts level---and fixed, but not until well after the completion of Milestone 2.  

The ground-loop in particular proved particularly troublesome, as it caused the bias voltage to fluctuate and effectively actuate the entire DM at once. If all of the actuators moved the same amount, this would have presented itself as a piston term and would not have been an issue; unfortunately, the nm/V gain does vary from actuator to actuator, and so the gain map is printed onto the wavefront by this noise.  Direct measurement of the voltage on the bias line showed the noise to be white with an RMS deviation of 0.035V.   It has been fixed, as verified both electrically (in an oscilloscope) and optically (in the dark-hole speckle field), and we do not expect this to be an issue in the future.

\subsection{Model comparisons}

As discussed previously, a closed-loop simulation with controllable static errors removes those errors efficiently and converges to very high contrasts at rates determined by the regularization parameter fed to the EFC algorithm.  Unfortunately, the dominant errors during Milestone 2, such as the electronic noise, were not static, and a model of the Milestone 2 testbed will be ineffective if it does not capture them.

Capturing the errors directly is not a trivial exercise.  Leaving aside DM temporal variability, we still have model mismatches, sensor noise, and slow drifts that limit the ability of a single iteration to reach high contrast. It is not possible to determine after the fact what all of these confounding factors were, and in some cases these may still be open areas of investigation.

We instead take a different tack: to capture the consequences of these variations, we vary the regularization parameter used in the model from iteration to iteration to match that of the testbed during a milestone run.  Each full EFC iteration consists of two subparts: a set of probe images, which provide data for the complex field estimation, and a set of regularization images, which test out different regularization parameters for a given correction.  The parameter which empirically provides the best performance given the testbed conditions at the time is selected to continue, implicitly minimizing the effect of the otherwise unknown errors in the system at that point in time.  Using that regularization parameter in the model for that iteration should thus best approximate the actual system state, in the absence of further knowledge of testbed internal state.

We also need to account for ``incoherent'' signal.  Pairwise estimation provides two pieces of data: a coherent, complex electric field estimate for the light that can be modulated by the deformable mirror, and an ``incoherent'' intensity estimate for the residual that does not modulate when the mirror is being used to probe.  This incoherent part can have many sources---stray light, estimation error, a planet (if used on the sky)---but regardless of the source, it is ignored in the subsequent estimation and correction. 

For our model comparison, we select the third milestone run, as it contains the fewest number of factors unrelated to correction.  The first quarter of run 1 contained several manual changes as we tested out the best control model, while run 2 was stopped and restarted briefly in the middle to recalculate the photometric correction factor.  

Our testbed model for this analysis uses a simple matrix-Fourier-based propagator\mycitep{Sou07} based in Matlab to move between pupil and image planes, and vice versa.  A more complex Fresnel model in PROPER\mycitep{Kri07} exists as well, with comparable results, but we have favored the Fourier model for speed; yet another Fourier-based model written in Python is used when the control is taking place on the testbed.  This model is also used to calculate the Jacobian for EFC correction.  The system is given a starting phase residual with 1.77nm rms error over controllable frequencies to initialize the contrast prior to correction.  

During the third testbed run (i.e. the data on the right side of Fig. \ref{fig:iters}), the control loop chose a new regularization parameter to maximize the contrast gain at each step.  This was done empirically, by taking images for 6 different DM settings.  As Fig. \ref{fig:cl1b} demonstrated, the rate of change in contrast varies strongly with the choice of the regularization parameter, so to replicate the performance in the model, we also use the same set of regularization parameters used for milestone run 3. After computing the mean coherent intensity across the dark hole, we add the mean incoherent background measured during that iteration.

\begin{figure}
\centering
\includegraphics[width=0.8\textwidth]{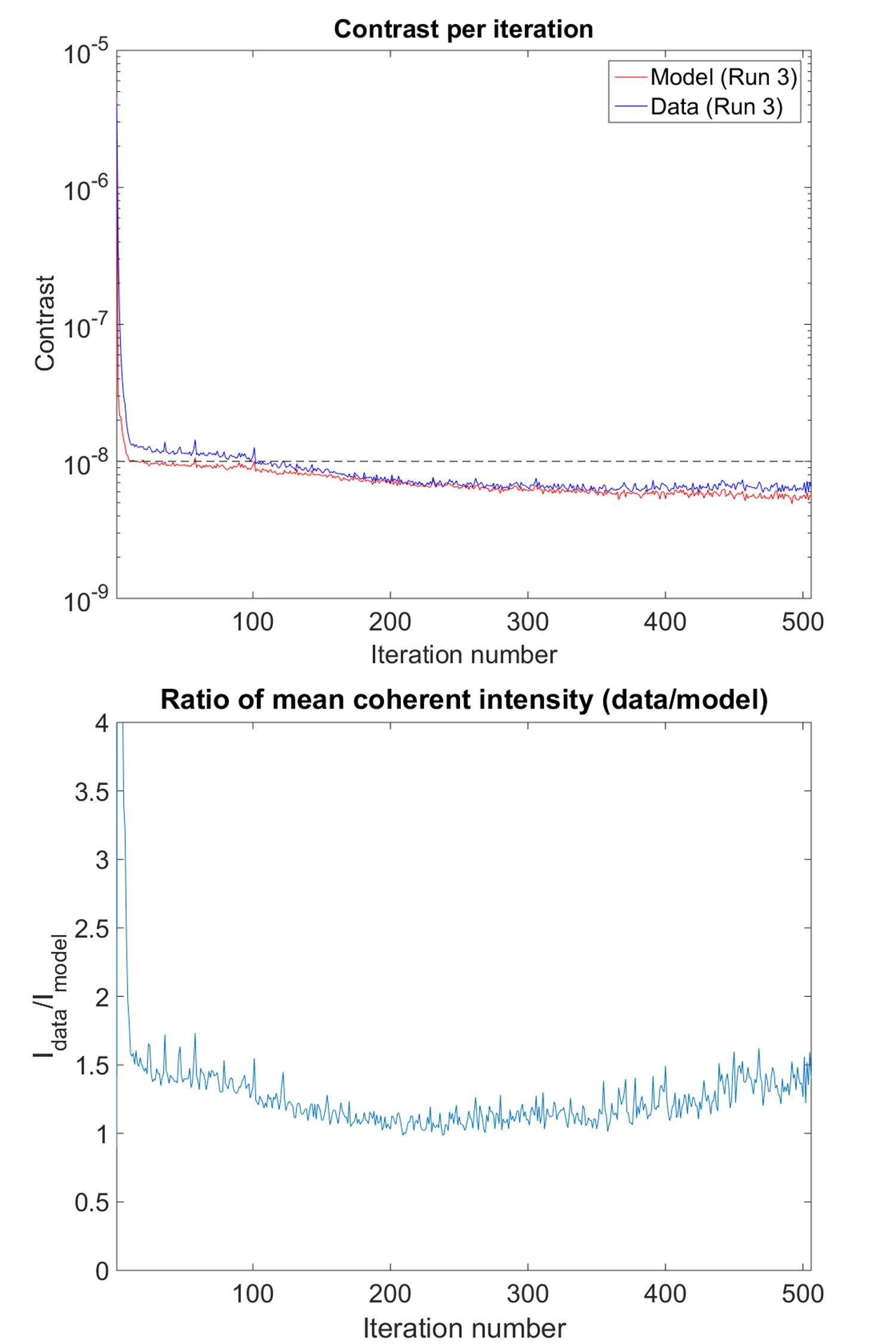}
\caption{\textit{Top:} Estimated mean contrast in the dark hole from milestone run 3, and a model of milestone run 3.  \textit{Bottom:}  Ratio of the coherent intensity fractions of the data and model, as a function of iteration.  (Including the incoherent part would make the model appear to be performing better it is.)  The peak at the far left reaches a maximum of 15.2.} \label{fig:mvsd}
\end{figure}

\begin{figure}
\centering
\includegraphics[width=1\textwidth]{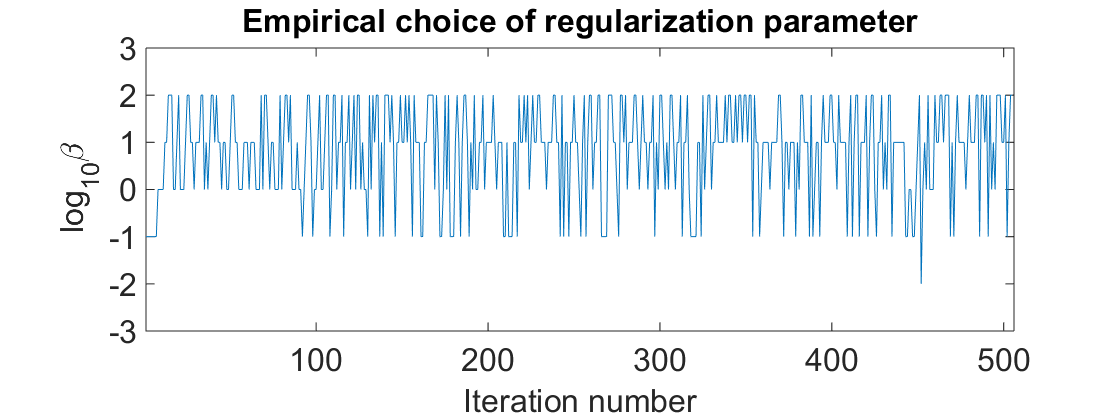}
\caption{Regularization parameter chosen in the testbed on each iteration of run 3.  Six values were tested during each iteration: $\beta = 10^{-3}, 10^{-2}, 10^{-1}, 10^{0}, 10^{1}, 10^{2}$.  The $\beta = 10^{-3}$ regularization was never the optimal choice during this run.  See Figure \ref{fig:cl1b} for the accompanying theoretical convergence rates.  Smaller $\beta$-values correspond to corrections with higher actuator stroke and more weight given to the model; generally, when the empirically-selected $\beta$ is smaller, the model used in the correction is a better representation of the true state of the testbed.} \label{fig:reglist}
\end{figure}

The top of Fig. \ref{fig:mvsd} shows the run 3 data overlaid with iterations from a model with incoherent background and regularization updates.  In the lower part, the ratio between the mean coherent intensity in the model and in the testbed are shown.  For the majority of the iteration sequence, this scheme captures the testbed performance to within a factor of two.  The notable exception occurs at the beginning of the control sequence, when the model converges significantly more quickly, peaking at a data-to-model ratio of 15.2.  During this period, the testbed is using some of its more aggressive regularization parameters ($\beta = 10^{-1}$, compared to $\beta = 10^{1}$ or even $=10^{2}$ later), which only reappear very intermittently later in the sequence. (The full list of regularization parameters is given in Figure \ref{fig:reglist}, and we attribute the high $\beta$ chosen by most iterations to the presence of the noise in the electronic bias level.) The cause for this discrepancy is unclear, though we suggest it is likely due to the internal testbed software model not capturing the alignment of coronagraph optics with sufficient accuracy.

\subsection{Planet yield}

\begin{figure}
\centering
\includegraphics[width=0.8\textwidth]{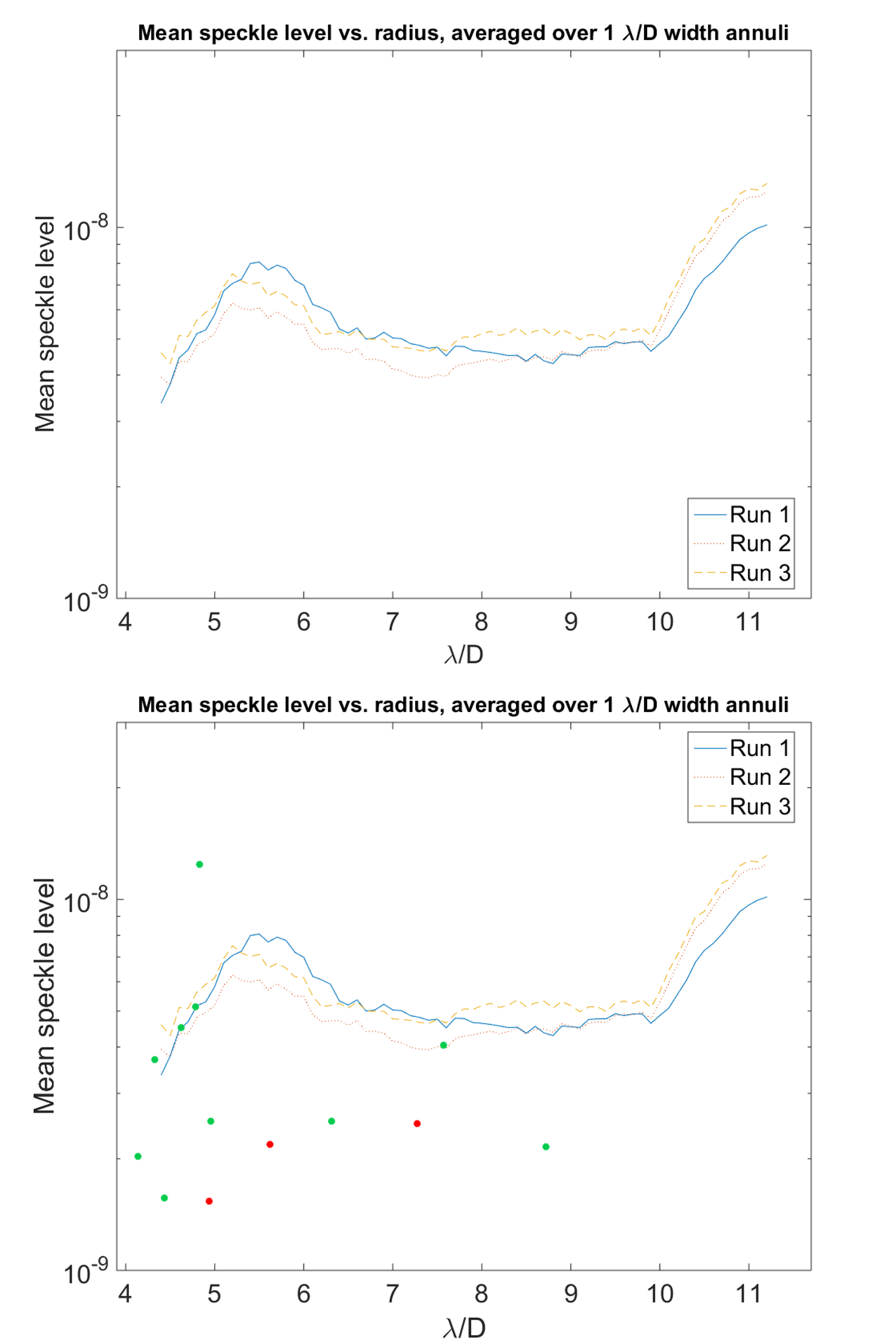}
\caption{\textit{Top.} Average radial contrast distribution for the three runs.  At each radial position, contrast is averaged over a $1 \lambda/D$ annulus centered at that radius. \textit{Bottom.} Distribution of known radial velocity planets with contrast curves from the three runs.  Planet size shows the expected flux at the IFS detector in a $2\%$ band.  Green planets are detectable with $10\times$ post-processing at SNR = 5; red planets are not detectable without $30\times$ post-processing.  One red planet falls below the x-axis of the plot and is not visible.}\label{fig:contrad}
\end{figure}

While planet-yield estimation currently relies on post-processing assumptions that are estimated but not yet validated, we can make some estimates to ensure that the contrasts we are demonstrating in the laboratory have some traceability to expected science with WFIRST-AFTA.  Working on the conservative WFIRST-AFTA assumption of $10\times$ reduction of speckle noise from post-processing, along with the catalog and planet yield assumptions such as albedo given in \mycitet{Tra14}, we expect 10 known-radial-velocity planets to be detectable with SNR of 5, assuming a $2\%$ band (i.e. looking through the IFS).  If $30\times$ gain from postprocessing is possible, we expect that number to rise to 14.  (Planet locations and radial contrast averages are shown in Figure \ref{fig:contrad}.) The next-generation SPLC is expected to have a higher yield than this, particularly thanks to its smaller IWA.

\subsection{Broadband performance} \label{subsec:bb}

\begin{figure}
\centering
\includegraphics[width=1\textwidth]{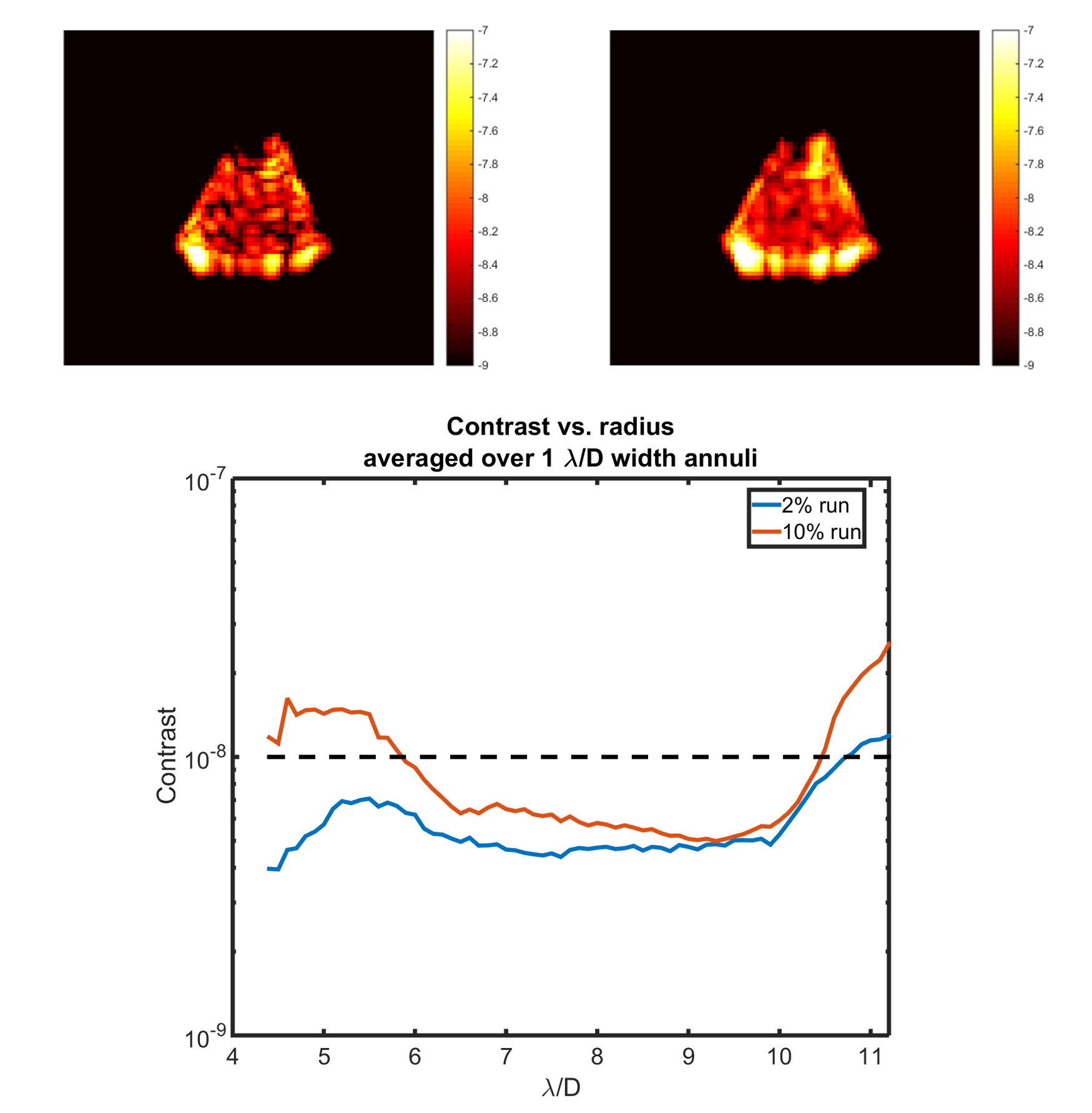}
\caption{\textit{Top left:} Dark hole in $2\%$ band centered at 550nm.  \textit{Top right:} Dark hole in $10\%$ band centered at 550nm, taken with same DM setting. The software dark-hole mask is not applied so that the PSF leaking into the dark hole at the inner and outer edges of the dark hole is more visible. Results are in units of contrast, on a log$_{10}$ scale. \textit{Bottom:} Azimuthally-averaged contrasts for the left and right images.
} \label{fig:bbcont}
\end{figure}

While Milestone 2 requirements did not specify broadband performance, we took advantage of the flexibility of our tunable filter to do a $2\%$ correction at 550nm and examine performance with a $10\%$ band also centered at 550nm.  (This is not the same as meeting Milestone 5 requirements early, since the architecture of choice for demonstrating that milestone is an SPLC.)  For this, we started correcting from the end point of run 2 to minimize number of iterations; it only took 2 iterations to bring the mean contrast in the dark hole back to $\sim 6 \times 10^{-9}$ even after the passage of several days.  Despite not correcting the ends of the band explicitly, the contrast remained quite low: $9.1\times 10^{-9}$ mean across the dark hole.  (An image of the data, photometrically corrected, is shown in Figure \ref{fig:bbcont}.)   

\section{Milestone 5 and beyond} \label{sec:fut}

Milestone 2 served as a basic functionality test for a reflective shaped pupil mask and an obscured aperture; Milestone 5 will act as a performance test to reach high contrast under more relevant instrument conditions.  Specifically, the Milestone 5 definition is
\begin{center}
\textit{The Occulting Mask Coronagraph in the High Contrast Imaging Testbed demonstrates $10^{-8}$ raw contrast with $10\%$ broadband light centered at 550 nm in a static environment.}
\end{center}
with the Occulting Mask Coronagraph being the catch-all term for the combination of the SPC and HLC in a single instrument.  While the demonstration in Sect. \ref{subsec:bb} is not done with the architecture we are specifying for Milestone 5---an SPLC rather than a first-generation SPC---it is a promising result in light of the Milestone 5 requirements.

Three major hardware/software changes have been implemented as part of the Milestone 5 run-up:
\begin{itemize}
\item A second DM was introduced 1m downstream of the first.  This permits two-sided control of amplitude and phase.
\item We have switched from measuring over a $2\%$-band to a $10\%$ band for milestone tests.  Control is done in a set of 5 $2\%$ sub-bands covering the broader $10\%$-band, which is both consistent with previous HCIT experience in broadband control and consistent with the primary use of the shaped pupil in flight: with an IFS whose channels will be $\leq2\%$ across.
\item Switch from a shaped pupil coronagraph to a shaped pupil Lyot coronagraph.  For this, all masks were swapped out for new designs, and the Lyot stop was installed downstream of the bowties.
\end{itemize}
All of these changes have been completed, and testing on the SPLC has begun.

We have demonstrated here that the shaped pupil coronagraph is capable of repeatedly reaching contrasts suitable for imaging planets with WFIRST-AFTA, with $\sim 6 \times 10^{-9}$ contrast averaged across the dark hole.  We have strong indications we will be able to continue to reach good contrasts in subsequent broadband tests for later milestones, and we have new designs expected to push this performance to better inner working angles and higher throughputs.

\section*{Acknowledgments}

The authors would like to thank H. Zhou, J. Jewell, and J. Krist for modeling support and assistance, D. Ryan for hardware support, K. Patterson for hardware design and fabrication, and our reviewers for a careful and thorough review of the paper.  The research was carried out at the Jet Propulsion Laboratory, California Institute of Technology, under a contract with the National Aeronautics and Space Administration. 


\listoffigures
\listoftables

\end{spacing}
\end{document}